\documentclass{aa}  

\usepackage{graphicx}
\usepackage{amsmath}
\usepackage{xspace}
\usepackage{subcaption} 
\usepackage{upgreek}
\usepackage{float}
\usepackage{mathabx}
\usepackage{hyperref}
\usepackage{tablefootnote}
\usepackage{tikz}
\usepackage{booktabs}
\usepackage{txfonts}

\newcommand\exo{EXO~2030+375\xspace}

\newcommand\nustar{\textsl{NuSTAR}\xspace}

\newcommand\exosat{\textsl{EXOSAT}\xspace}
\newcommand\rxte{\textsl{RXTE}\xspace}
\newcommand\isgri{\textsl{INTEGRAL}-ISGRI\xspace}
\newcommand\nicer{\textsl{NICER}\xspace}
\newcommand\maxi{\textsl{MAXI}\xspace}
\newcommand\ixpe{\textsl{IXPE}\xspace}
\newcommand\hxmt{\textsl{Insight}-HXMT\xspace}

\newcommand\chandra{\textsl{Chandra}\xspace}
\newcommand\integral{\textsl{INTEGRAL}\xspace}

\newcommand\swift{\textsl{Swift}\xspace}

\bibpunct{(}{)}{;}{a}{}{,} %

\begin{document} 

   \title{The giant outburst of EXO~2030+375}
   \subtitle{I: Spectral and pulse profile evolution}
   \titlerunning{Giant outburst of EXO 2030+375}

   \author{P. Thalhammer\inst{\ref{in:rem}}
          \and R.~Ballhausen\inst{\ref{in:UMD},\ref{in:gsfc}}
          \and E.~Sokolova-Lapa\inst{\ref{in:rem}}
          \and J.~Stierhof\inst{\ref{in:rem}}
          \and A.~Zainab\inst{\ref{in:rem}}
          \and R.~Staubert\inst{\ref{in:UT}}  
          \and K.~Pottschmidt\inst{\ref{in:UMBC},\ref{in:gsfc}}
          \and J.~B.~Coley\inst{\ref{in:howard},\ref{in:gsfc}}
          \and R.~E.~Rothschild\inst{\ref{in:UCSD}}
          \and G.~K.~Jaisawal\inst{\ref{in:NSI}}
          \and B.~West\inst{\ref{in:USNA}}
          \and P.~A.~Becker\inst{\ref{in:GMU}}
          \and P.~Pradhan\inst{7,8}
          \and P.~Kretschmar\inst{9}
          \and J.~Wilms\inst{\ref{in:rem}}
        }

        \institute{
        {Dr.\ Karl Remeis-Observatory and Erlangen Centre
          for Astroparticle Physics,
            Friedrich-Alexander Universit\"at Erlangen-N\"urnberg, Sternwartstr.~7, 96049
            Bamberg, Germany \label{in:rem}}
         \and
          {University of Maryland College Park, Department of Astronomy, College Park, MD 20742, USA \label{in:UMD}}
          \and
           {University  of Maryland Baltimore County, 1000 Hilltop Circle,
          Baltimore, MD 21250, USA\label{in:UMBC}}
          \and
          {CRESST and NASA Goddard Space Flight Center, Astrophysics Science Division, 8800 Greenbelt Road, Greenbelt, MD
          20771, USA\label{in:gsfc}}
          \and
          {Institut für Astronomie und Astrophysik, Universität Tübingen, Sand 1, 72076 Tübingen, Germany \label{in:UT}  } 
          \and
          {Department of Physics and Astronomy, Howard University, Washington, DC 20059, USA \label{in:howard}}
          \and 
          Massachusetts Institute of Technology, Kavli Institute for Astrophysics and Space Research, 70 Vassar St., Cambridge, MA 02139 
          \and 
          Department of Physics and Astronomy, Embry-Riddle Aeronautical University, 3700 Willow Creek Road, Prescott, AZ 86301, USA
          \and
          European Space Agency (ESA), European Space Astronomy Centre (ESAC), Camino Bajo del Castillo s/n, 28692 Villanueva de la Cañada, Madrid, Spain
          \and 
          {Department of Astronomy and Astrophysics, University of California, San Diego, 9500 Gilman Dr., La Jolla, CA 92093-0424, USA \label{in:UCSD}}
          \and 
          {Department of Physics and Astronomy, George Mason University, Fairfax, VA 22030-4444, USA\label{in:GMU}}
          \and 
          {National Space Institute, Technical University of Denmark, Elektrovej 327-328, DK-2800 Lyngby, Denmark\label{in:NSI}}
          \and 
          {Department of Physics, United States Naval Academy, Annapolis, MD 21402, USA \label{in:USNA}}
        }

        \date{RECEIVED: ACCEPTED:}

  \abstract{ 
The Be X-ray binary \exo went through its third recorded giant outburst from June
2021 to early 2022. We present the results of both spectral and timing analysis
based on \nicer monitoring, covering the 2--10\,keV flux range from 20 to 310\,mCrab. 
Dense monitoring with observations carried out
about every second day and a total exposure time of $\sim$160\,ks allowed us to
closely track the source evolution over the outburst. Changes in the spectral shape
and pulse profiles showed a stable luminosity dependence during the rise and
decline. The same type of dependence has been seen in past outbursts. The pulse
profile is characterized by several distinct peaks and dips. The profiles show a clear dependence on luminosity with a stark transition at a luminosity of ${\sim}2\times10^{36} \,\mathrm{erg\,s}^{-1}$, indicating a change in the emission pattern. Using relativistic raytracing, we demonstrate how anisotropic beaming of emission from an accretion channel with a constant geometrical configuration can give rise to the observed pulse profiles over a range of luminosities.
}


\maketitle
\section{Introduction}

Accreting Be X-ray binary systems (BeXRBs) are a common target for the study of accretion onto highly magnetized neutron stars as they have the advantage of being both common and going through a wide range of luminosities during an outburst \citep{reig_2011,Okazaki_2013,Vybornov_2017}. 
Depending on the mass-accretion rate, these systems feature discrete luminosity states associated with changing properties of the X-ray emission regions in the accretion channel closer to the surface of the neutron star, the ``accretion column.'' These changes are thought to be related to the geometry of the accretion column due to the different physical mechanisms with which the accreted matter is decelerated from its free fall speed of ${\sim}0.5c$ \citep[see, e.g.,][]{becker_2012}.
Generally, two distinct types of outbursts are observed in BeXRBs: regular Type~I outbursts that occur close to periastron and last for about 20--50\% of the system's orbital period, and giant Type~II outbursts. The latter ones can span up to two orders of magnitude in luminosity. 
 
\begin{figure}\label{fig:lc_exo}
  \resizebox{\hsize}{!}{\includegraphics{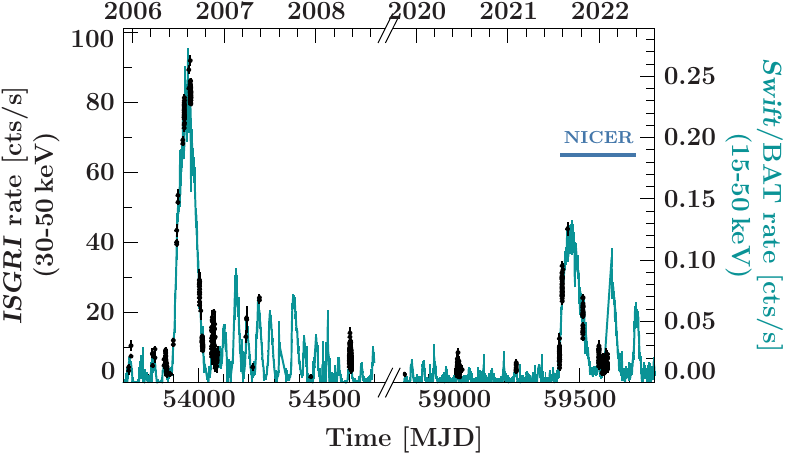}} 
  \caption{Long-term light curve of \exo in the 30--50\,keV band as seen with \isgri in black and the \swift-BAT light curve in teal. For comparison, time ranges around the two most recent Type~II outbursts are shown.  \isgri data were binned by Science Window, while for \swift-BAT the daily light curve was used. Marked by a horizontal bar is also the time range covered by the \nicer monitoring used in this work. See Fig.~\ref{fig:lc} for the \nicer light curve of the 2021 outburst.}\label{fig:isgri}
\end{figure}

One prototypical system that shows both types of outbursts very prominently is \exo (see Fig.~\ref{fig:isgri}), a neutron star with a B0 Ve companion \citep{reig_timing_1998}. The source was discovered with  European X-ray Observatory Satellite (\exosat) in 1985 during a Type~II outburst showing clear pulsations with a period of 43\,s \citep{parmar_transient_1989}. 
These early observations also showed orbital modulation of the pulse period, establishing its 46\,d orbit \citep{wilson_outbursts_2008}. This orbital motion results in regular Type~I outbursts around the periastron passage \citep{stollberg_recent_1994}. The luminosity of Type~I outbursts and phase shift with respect to the periastron varies with time and is influenced by the preceding giant outburst, possibly due to changes in the circumstellar disk \citep{laplace_possible_2017}.  Since its initial discovery, \exo has continued to display regular Type~I outbursts and pulsations down to 3--38\,keV luminosities of ${\sim}8\times 10^{34}\,\mathrm{erg}\,\mathrm{s}^{-1}$ \citep{furst_studying_2017}.

The early \exosat observation already covered a factor of $\sim$100 in luminosity and revealed a stark variability of pulse profiles with luminosity. At  1--20\,keV luminosities upward of ${\sim}2.4\times 10^{37}\,\mathrm{erg}\,\mathrm{s}^{-1}$, the profiles are dominated by a single broad peak, which initially becomes sharper and more asymmetric with lower luminosity before a secondary peak becomes dominant when reaching a luminosity of ${\sim}2.8\times 10^{35}\,\mathrm{erg}\,\mathrm{s}^{-1}$. Additionally, as the luminosity decreases to around ${\sim}2.8\times 10^{36}\,\mathrm{erg}\,\mathrm{s}^{-1}$, a distinct ``notch'' forms in the profile \citep{parmar_transient_1989-1}\footnote{We note that the distance of the companion of \exo derived from  Gaia Data Release 3 (DR3) is $2.4_{-0.4}^{+0.5}\,$kpc \citep{Bailer-Jones_2021}. This distance is significantly shorter than that previously assumed  \citep[$7.1\pm0.2\,$kpc][]{wilson_decade_2002}. As such, most of the literature so far has overestimated the distance to \exo and thus its luminosity. See Appendix~\ref{sec:distance} for a discussion. For this work, unless stated otherwise, we consistently give luminosities based on the updated Gaia DR3 distance.}.
This pattern of variability with luminosity has been repeated with little variation by all Type~I and Type~II outbursts \citep{wilson_outbursts_2008,epili_decade_2017}. 
Consequently, observations by the Rossi X-ray Timing Explorer (\rxte) during one of the Type~I outbursts showed no significant change in the pulse profiles compared to previous \exosat observations at a comparable luminosity, indicating the remarkable consistency of the luminosity-dependent pulse profiles in \exo \citep{wilson_outbursts_2008}.
A second giant Type~II outburst was detected in 2006 \citep{klochkov_integral_2007}, again showing the familiar dependence on the luminosity of the pulse profile. 
Concerning the energy dependence of the profiles \citet{parmar_transient_1989-1} initially noted a resemblance of the high-energy profiles to those at lower luminosities. Subsequent phase-resolved spectral analysis revealed significant changes in the parameters of the spectral continuum with the pulse phase \citep{klochkov_giant_2008}.
In July 2021, a third Type~II outburst was detected. This outburst was monitored by multiple X-ray missions, including, the Neutron Star Interior Composition Explorer (\nicer), the Nuclear Spectroscopic Telescope Array (\nustar), \swift, \chandra, and the International Gamma-Ray Astrophysics Laboratory (\integral), among others.
The goal of this paper is to make use of this unprecedented data coverage on the complex but reproducible pulse profile behavior of \exo and to provide insight into possible emission geometries by expanding on early modeling attempts.
Although some of the data presented here have been published by \citet{tamang_spectral_2022}, these authors only briefly touched on the evolution of the pulse profile and its physical interpretation.  Here, we aim to provide a more detailed analysis in the following. We begin our paper with an overview of the available \nustar and \nicer observations and the respective data extraction, continuing with a brief look at the spectral evolution of \exo over its outburst in Sect.~\ref{sec:nispec}. The main part of the paper focuses on the evolution of the pulse profile, starting with an overview of the \nicer profile shape evolution with the luminosity and energy seen with \nustar in Sect.~\ref{sec:map}. In Sect.~\ref{sec:mod} we explain how we applied the physical light-bending code to our two specific observations and expanded our pulse profile analysis to all observations through the use of a simplified phenomenological model in Sect.~\ref{sec:fit}. Finally, we discuss our results in Sect.~\ref{sec:dis}.
An accompanying paper, Ballhausen et al. (\citet{Ballhausen_2023}, hereafter Paper II), will focus on broadband spectroscopy at the outburst peak and at the end of the outburst with \nicer and \nustar.

\section{Observations and data reduction} \label{sec:obs}

\begin{figure*}
  \resizebox{\hsize}{!}{\includegraphics{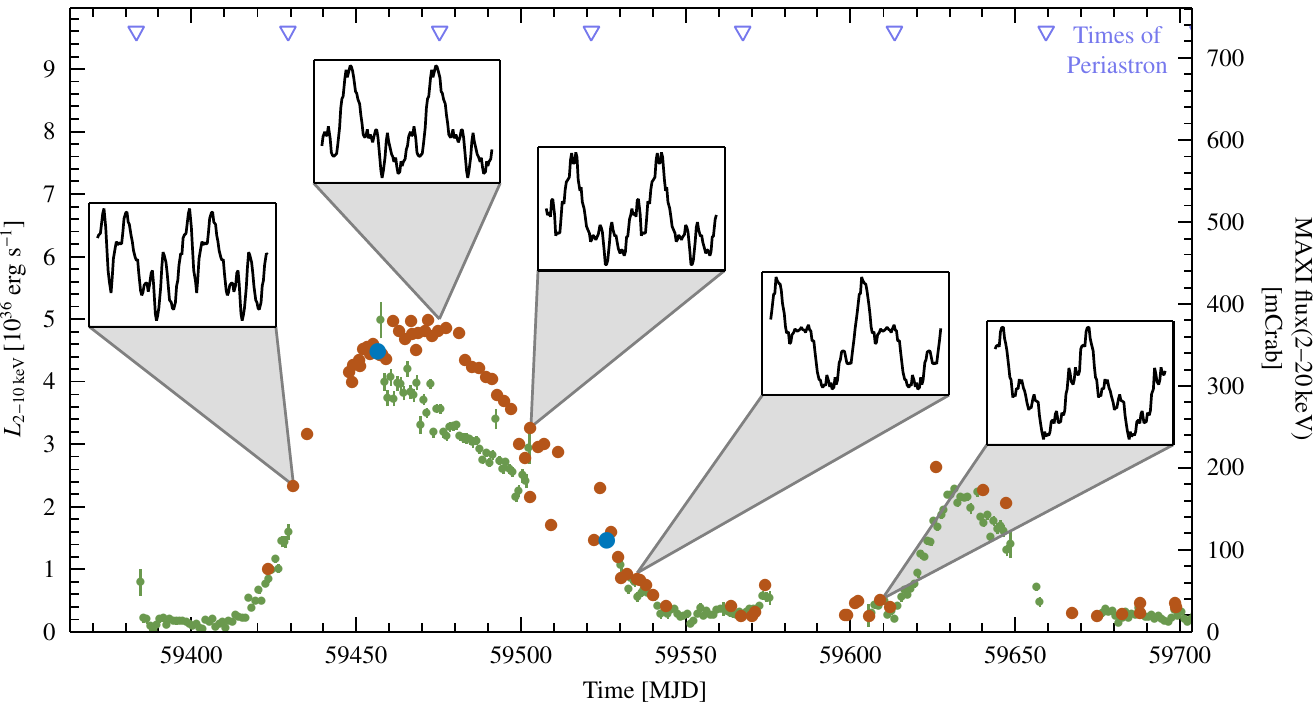}}
  \caption{Light curve of the outburst as seen by \nicer with each point corresponding to one observation. The luminosity of each observation is derived from the spectral fits described in Sect.~\ref{sec:nispec}. The daily \maxi light-curve of the same period is shown in green. Observations with quasi-simultaneous \nustar points are marked in blue. Overlaid are five of the pulse profiles at different times over the outburst. The purple markers at the top indicate the times of periastron passage using the orbital solution by \citet{wilson_decade_2002}.}\label{fig:lc}
\end{figure*}

\subsection{NICER} \label{sec:nic}

Figure~\ref{fig:isgri} shows the long-term light curve of \exo in the 20--80\,keV band measured with the INTEGRAL Soft Gamma-Ray Imager (ISGRI)  and \swift-BAT. The figure shows the 2006 Type~II giant outburst, which was followed by a sequence of lower flux Type~I outbursts with a decaying flux. In 2021, the start of a new Type~II outburst was detected by the Monitor of All-sky X-ray Image (\maxi) \citep{akajima_atel_2022}. We triggered \nicer monitoring of the outburst, which began monitoring \exo on 2021 August 5, and continued with regular observations of decreasing cadence until at least 2023 March (${\sim}$ MJD 60016). For this work, we include all observations until ObsID 4201960176 (MJD~59609), amounting to and exposure time of  ${\sim}160\,$ks. Figure~\ref{fig:lc} shows the resulting \nicer light curve over the course of the outburst. Some gaps in the monitoring during the rise and decline of the outburst are an inevitable result of the visibility constraints that come with \nicer's location on board the International Space Station (ISS). 
We extracted the data using the \nicer data analysis pipeline, as part of Heasoft 6.31.1 with CALDB version 20220413. 

For data reduction, we used the default filter criteria used by \textit{nicerl2}. Background spectra were generated using the \textit{3C50} model\footnote{\url{https://heasarc.gsfc.nasa.gov/docs/nicer/analysis_threads/background/}}. We explored the use of alternative background models, but found the  \textit{3C50} to produce the most consistent background spectra.  We further note that the source was bright enough that the choice of background model did not affect the results of our spectral analysis in a significant way. Unless otherwise stated, we restricted our spectral analysis to the 0.5--10\,keV energy band, where the \nicer calibration is most reliable.

\subsection{\nustar} \label{sec:nu}
In addition to the \nicer monitoring, we also performed two \nustar observations. The first observation was performed on 2021 August~30, when it became apparent that the outburst approached its peak, and the second on 2021 November~8 during the decline, when \nicer began to detect significant changes in the pulse profile. These two observations, with ObsID 80701320002 and ObsID 90701336002, have exposure times of 32\,ks and 23\,ks, respectively. Light curves for various energy bands were generated using version 0.4.9 of the \textit{nupipeline} and CALDB v20211020. Pulse profiles were constructed from barycentered light curves corrected for dead time with 0.5\,s resolution. See the accompanying Paper~II for a detailed discussion of these two observations. A detailed description of the data reduction is given in Paper II.

\section{Spectral evolution over the outburst} \label{sec:nispec}
\begin{figure}
  \resizebox{\hsize}{!}{\includegraphics{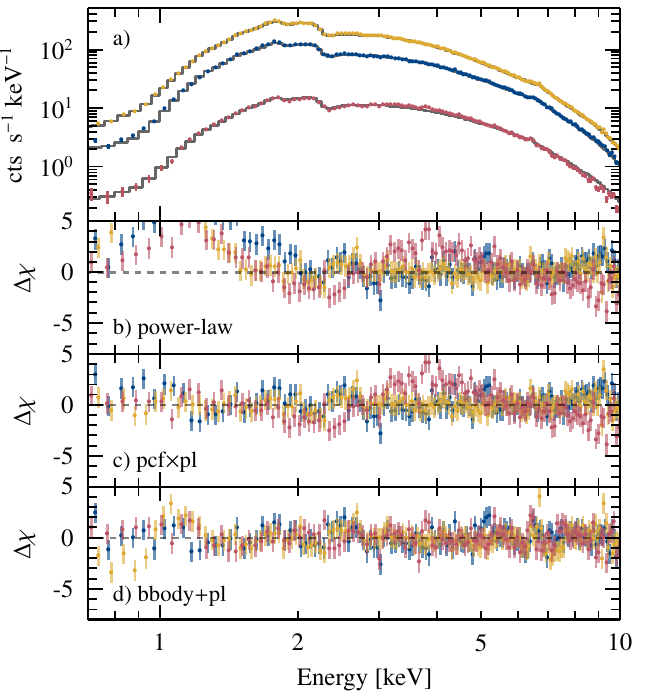}}
  \caption{Spectrum of the rise (ObsID 4201960102), peak (ObsID 4201960113), and decline (ObsID 4201960150) of the outburst of \exo  and their best fit with the \texttt{bbody+pl} model in blue, yellow, and red respectively. The lower panels show the residuals for fits with an absorbed power law with an iron K$\alpha$ line emission and with an additional black body or partial covering component. } \label{fig:spec}
\end{figure}

\renewcommand{\arraystretch}{1.5}
\begin{table}
     \caption{Mean and standard deviation of reduced $\chi^2$ values for the best fits of the three considered models to the \nicer observations studied in this work. }
    \centering
    \begin{tabular}{c|ccc}
    \toprule
         & \texttt{pow} & \texttt{pow+pcf} & \texttt{pow+bb} \\
         \hline
        mean red. $\chi^2$      &  $2.31$  &  $1.61$ &  $ 1.02$ \\
        stddev. red. $\chi^2$ &  $0.95$  &  $0.91$ &  $ 0.35$ \\
            \hline

    \end{tabular}
    \label{tab:fitchi}
\end{table}

\begin{figure*}
\sidecaption
\includegraphics[width=120mm]{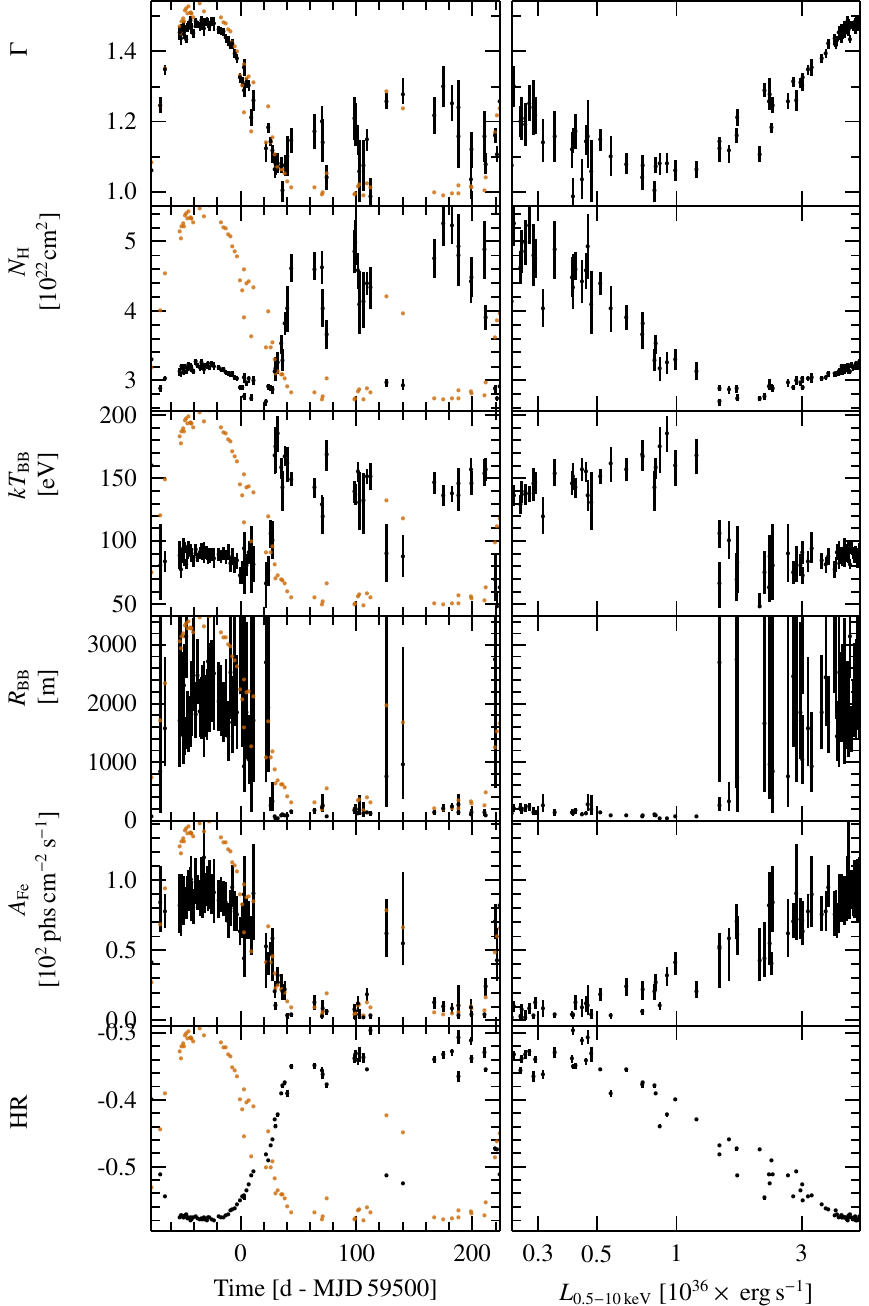}
  \caption{Evolution of the 0.5--10\,keV flux, hardness, and main spectral parameters over the course of the 2021 outburst. The left panel shows the evolution over time and the right panel shows how these parameters depend on luminosity. The orange data points in the left panel represent the light curve over the course of the outburst. The hardness is defined as $(H-S)/(H+S)$ where $H$ is the count rate in the 0.5--2\,keV band and where $S$ is the count rate in the 2--10 keV band. Each point corresponds to one observation. }\label{fig:spec_evo}
\end{figure*}

To put the evolution of the pulse profile in the context of the spectral evolution over the course of the outburst, we describe the \nicer spectra with an empirical continuum model. A continuum model consisting of an absorbed power law with a Gaussian iron line, using abundances by \citet{wilms_2000} and cross sections by \citet{Verner_1996}, shows significant residuals in the soft X-rays. We addressed those in two ways, either with an additional blackbody component, as used among others by \citet{reynolds_luminosity_1993}, or with an additional partial covering absorber as also applied before to this source by \citet{naik_suzaku_2015}.
A comparison of residuals between the three models during the rise, peak, and decline of the outburst is shown in Fig.~\ref{fig:spec}. Both the added blackbody and the partial coverer\footnote{https://pulsar.sternwarte.uni-erlangen.de/wilms/research/tbabs/} lead to a significantly improved description of the data with a good description of the soft excess below ${\sim}1.4$\,keV. The remaining residuals around 2\,keV are likely the result of calibration uncertainties around the Au-edge in the \nicer effective area \citep{Vivekanand_2021}.

To track the spectral change over the outburst we chose to model the soft excess around 1\,keV with a black body emission component, as this led to slightly
improved fit statistics compared to the partial coverer. However, we caution that the additional soft emission components cannot be fully disentangled from a geometrically complex and potentially ionized absorber and that the spectrum is likely affected by both. 
An overview of the average fit statistic for the three tested models is shown in Tab.~\ref{tab:fitchi}. 
The resulting evolution of the spectral parameters is shown in Fig.~\ref{fig:spec_evo}. The panel on the right displays the correlation of spectral parameters with luminosity. 
Both the hardness ratio and the photon index show a clear trend toward a softer spectrum with increased flux, while the fitted hydrogen column density switches from a negative to a positive correlation as luminosity increases.  There is a clear jump in the temperature and size of the soft component around a luminosity of $10^{36}\,\mathrm{erg}\,\mathrm{s}^{-1}$, which also coincides with an inversion in the evolution of the photon index, $\Gamma$, and the absorption column, $N_\mathrm{H}$. The source reached this luminosity close to the end of the outburst (Fig.~\ref{fig:lc}). While one could interpret the change in the blackbody component as a transition from accretion disk emission to emission coming from the neutron star surface, we repeat that the soft spectrum is likely more complex than modeled here and should therefore be interpreted with caution. 

\begin{figure}
  \resizebox{\hsize}{!}{\includegraphics{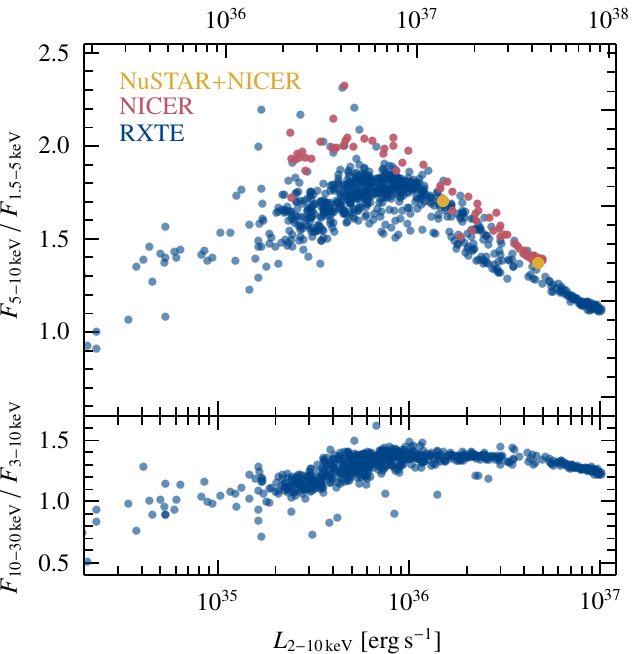}}
  \caption{Evolution of hardness, as derived from the flux ratio in the 1.5--5\,keV and 5--10\,keV bands. For the 2021 outburst, \nicer data was used, while for the 2006 outburst we used archival \textit{RXTE} data. For comparison with previous works, a distance of 7.1\,kpc was used to calculate the luminosity on the upper x-axis, while 2.4\,kpc was used for the lower x-axis.}\label{fig:hardness}
\end{figure}

Figure~\ref{fig:hardness} puts the spectral behavior of \exo during the 2021 Type~II outburst in the context of the  Type~II outburst  observed in 2006 with \rxte. The \rxte data of \exo have been studied by various authors but here we focus on the evolution of spectral hardness with luminosity as presented by \citet{epili_decade_2017}. These authors chose energy bands of 3--10\,keV and 10--30\,keV for measuring the hardness ratio and source luminosity. Du to the hard band lying outside the \nicer bandpass and the rather complex evolution of the spectral shape of \exo, their results are complementary but not directly comparable to the behavior we observe. From the 2006 \rxte data we, therefore, construct hardness ratios and luminosity estimates using the same energy bands as for our analysis of the \nicer data. Additionally, we show in the lower panel of Fig. \ref{fig:comp} the hardness ratio as defined by \citet{epili_decade_2017}.  To determine the fluxes, we used the same spectral model that we also used for the 2021 data, with the exception of the blackbody components, which are not required in the \rxte/PCA energy range of 2--60\,keV. Using consistent energy bands and estimated distances, we find that the overall dependence of the spectral hardness on the X-ray luminosity is very similar between the 2006 and 2021 giant outbursts, although the \rxte data span a larger luminosity range overall and the 2006 outburst had a higher peak luminosity. While we note subtle differences in the shape of the turnover in comparison to \citet{epili_decade_2017} due to the different energy bands, the transition in spectral behavior around $L_{2\mbox{--}10\,\mathrm{keV}}\sim 10^{36}\,\mathrm{erg}\,\mathrm{s}^{-1}$ from a positive to a negative correlation between luminosity and hardness is consistent and strong evidence for a transition toward supercritical accretion \citep{postnov_dependence_2015}. 

\section{The varying pulse profile of \exo}\label{sec:map}
\subsection{Pulse-profile evolution with luminosity} \label{sec:map_lum}

\begin{figure*}
\centering
\includegraphics[width=17cm]{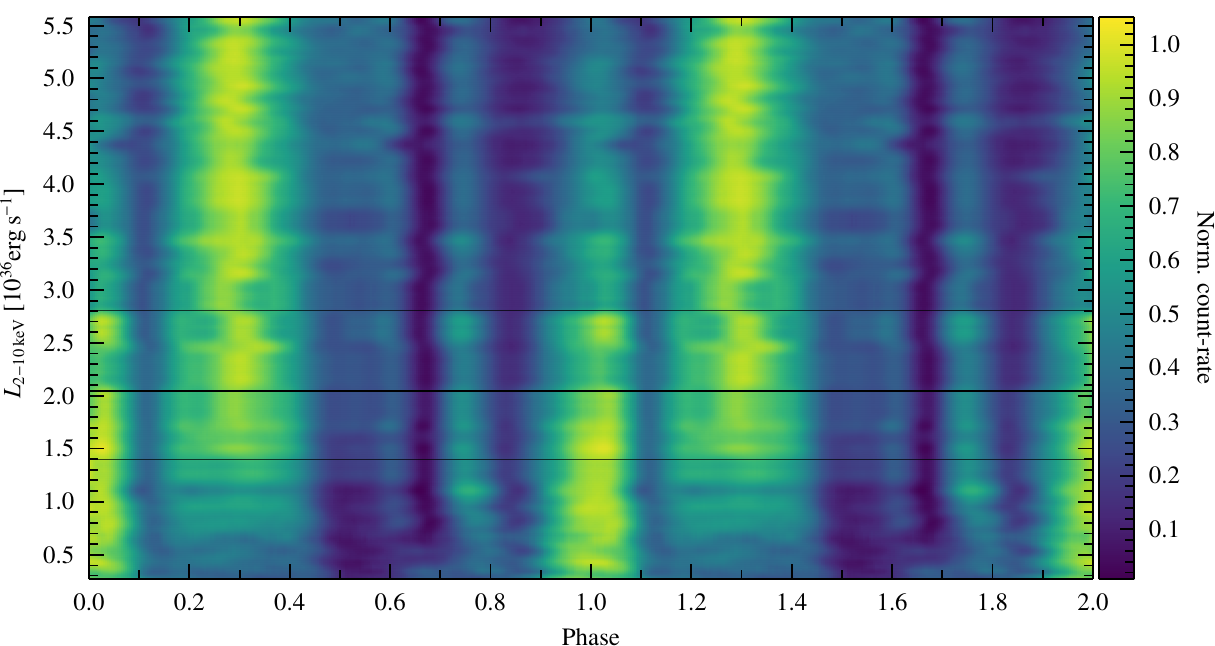}
  \caption{Pulse profile map scaled to range from zero to one for all included \nicer observations sorted by their respective luminosity. This figure illustrates the luminosity dependence of the \nicer pulse profiles. 
  The thick black line indicates luminosities where the profile shows marked changes, and the thin lines serve as markers to indicate the approximate luminosity range over which the changes take place (see the Sect. \ref{sec:map_lum} for more information).}\label{fig:map}
\end{figure*}

In addition to systematic spectral changes with luminosity, \exo is known for similarly remarkable changes in its pulse profile.  
Due to the strong spin-up of up to ${\sim}5\times10^{-8}\,\mathrm{s}\,\mathrm{s}^{-1}$\footnote{https://gammaray.nsstc.nasa.gov/gbm/science/pulsars.html}, leading to the period decreasing from 41.29\,s to 41.22\,s over the course of the outburst, we determined the local pulse period for each observation by epoch folding the orbit-corrected light curve and aligning the profiles through cross-correlation with respect to the profile of the preceding observation. To study how the average pulse profile changes with luminosity, we normalize the pulse profile for each observation in count rate, such that the peak count rate equals 1. Color coding of this normalized count rate allows us to compare the different pulse profiles in a pulse profile ``map,'' in which each row represents the profile at a given average count rate of the observation. This map is shown in Fig.~\ref{fig:map}. We note that the map contains profiles from both the increase of the outburst and the decrease of the outburst (although due to our observational cadence, the outburst decay dominates). We did not see a difference in profile between the increase and decay, indicating that the change of profile depends only on the source luminosity. As shown in Fig.~\ref{fig:map}, as the source luminosity increases, there is a clear change in strength between the first peak at phase 0 and the second around phase 0.25. In contrast, the overall structure of the profile between phases 0.5 and 0.9 is comparably stable. There also appears to be little further change in the profile above a count-rate of ${\sim}400$\,cps, corresponding to a luminosity of ${\sim}3\times10^{36}\,\mathrm{erg}\,\mathrm{s}^{-1}$. A very stable feature of the profiles is also the dip around phase 0.7, which only seems to lose its shape toward the lowest luminosities, where low statistics limits the usable phase resolution. 

\begin{figure*}
  \resizebox{\hsize}{!}{\includegraphics{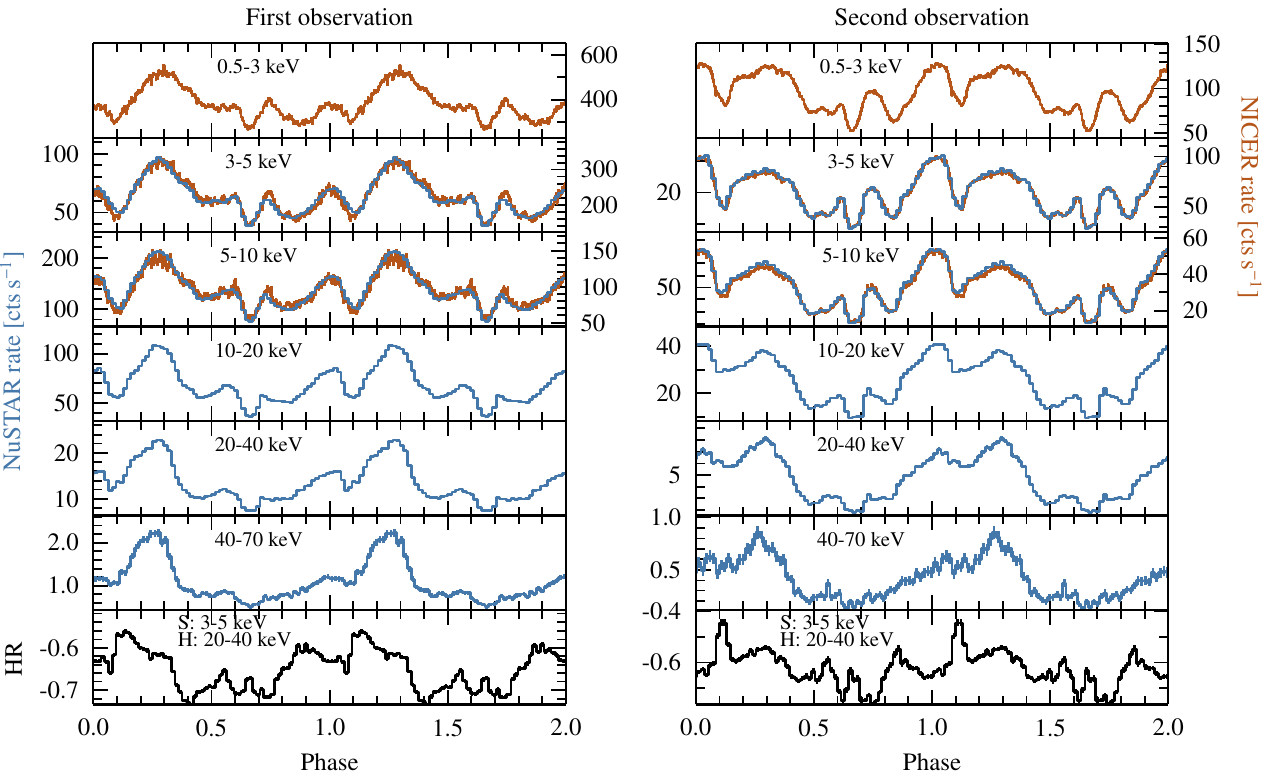}}
  \caption{Pulse profiles in different energy bands for both \nicer in blue and \nustar in red. The left plot shows the first \nustar observation and the concurrent  \nicer pointing, while the right plot shows the second \nustar observation and its concurrent \nicer pointing. The hardness ratio in the lowest panel is calculated from \nustar data as $(H+S)/(H-S)$, with the two energy bands S: 3--5\,keV and H: 20--40\,keV.    }\label{fig:comp}
\end{figure*}

\begin{figure}
  \resizebox{\hsize}{!}{\includegraphics{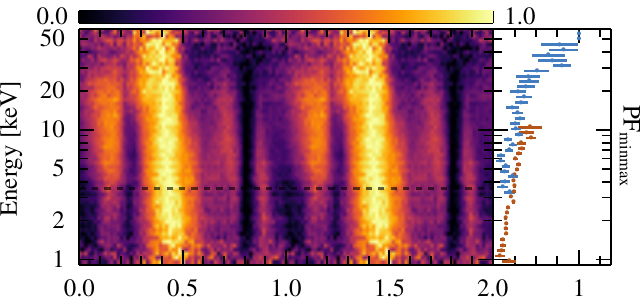}}
  \resizebox{\hsize}{!}{\includegraphics{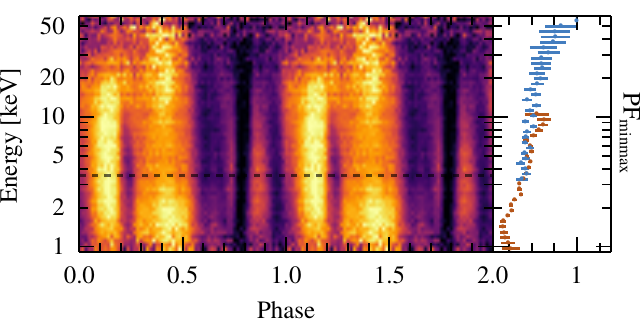}}
  \caption{Pulse profile map illustrating the energy dependence of the pulse profiles. The upper plot corresponds to the first  \nustar observation and the corresponding \nicer data, similarly for the second observation. The same data is used as for Fig.~\ref{fig:comp}.
  Dashed, black lines indicate the transition between \nicer and \nustar data. To the right, the pulsed fraction (PF) is plotted as a function of energy; blue is for \nustar data and in orange is for \nicer. We define the PF as $\mathrm{PF} = \left(p_\mathrm{max}-p_\mathrm{min}\right) / \left(p_\mathrm{max}+p_\mathrm{min}\right).$ }\label{fig:enmap}
\end{figure}

In addition to these changes with luminosity, the profile also depends strongly on energy.  To extend the energy range to harder photons, we include the NuSTAR data in this study. Consequently, we concentrate on the pulse profile during the peak of the outburst and during the end of the outburst, at X-ray luminosities of ${\sim}1.6\times10^{36}\,\mathrm{erg}\,\mathrm{s}^{-1}$ and ${\sim}4.78\times10^{36}\,\mathrm{erg}\,\mathrm{s}^{-1}$, respectively.  Figure~\ref{fig:comp} shows the pulse profiles for selected energy bands and the hardness ratio as a function of the pulse phase. The corresponding complete phase-energy maps of the same observations are shown in Fig.~\ref{fig:enmap}. Here, only strictly overlapping GTIs are used to keep especially the pulsed fraction between instruments comparable.  We find an evolution of the first peak at phase 0 and the wings around the dip at phase 0.7, which both seem to flatten toward the highest energies, leading to a simpler profile that is dominated by a single component. This trend to a less structured profile is present in both the high- and low-flux observations. 

Interestingly, when looking at the hardness ratio in Fig.~\ref{fig:comp}, around phase $\phi=0.07\text{--}0.15$  there seems to be a significant difference between the two observations. This phase corresponds to the decline of the first peak and appears significantly softer than surrounding phase bins during the first observations and harder in the second observation. This is discussed in more detail in Paper~II, where pulse phase-resolved spectroscopy reveals changes in the blackbody emission around this phase.

\subsection{Sharp dip at low luminosity}\label{sec:dip} 

A feature in the pulse profile that is only visible during observations at low luminosities of ${\sim} 2.0 \times 10^{35}\,\mathrm{erg}\,\mathrm{s}^{-1}$ is a sharp dip at phase 0.75. As such, it is only detected in some of the last observations, most prominently for ObsID 4201960165, which is shown in Fig. 9. The dip only spans a phase interval of 0.05 and is not resolvable in observations with low statistics. It is accompanied by a sharp change in the hardness ratio, marked by a hardening during the center of the peak and a softening during its flanks. The feature has been observed in several previous observations and has been attributed to absorption by a phase-locked accretion stream \citep{reig_x-ray_1999,naik_suzaku_2015,ferrigno_glancing_2016,furst_studying_2017,jaisawal_detection_2021}. The complex evolution of the pulse profile might be driven by a complex ionization profile of the absorbing medium.

\begin{figure}
  \resizebox{\hsize}{!}{\includegraphics{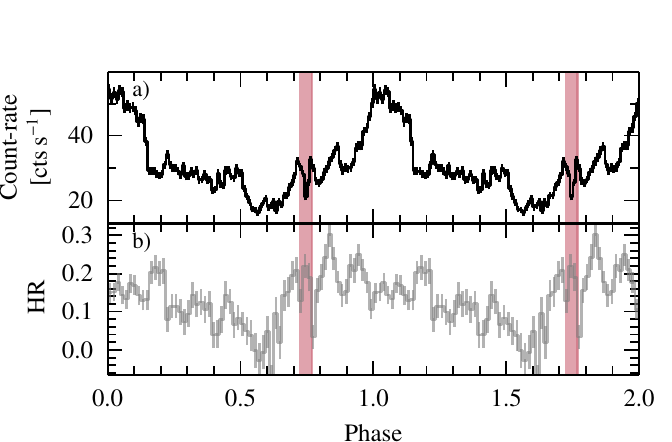}}
  \caption{Sharp dip around phase 0.75 during ObsID 4201960165 at a luminosity of $2.56\times10^{35} \mathrm{erg}\,\mathrm{s}^{-1}$, taken at MJD\,59570. The upper panel shows the count-rate in the full 0.5--10\,keV band, while the lower one shows the change in hardness ratio defined as in Fig.~\ref{fig:comp}, with 0.5--2.0\,keV  and 2.0--10.0\,keV as soft and hard energy bands. }\label{fig:dip}
\end{figure}

\section{Modeling the pulse profile of \exo}

\subsection{Physical pulse profile modeling}\label{sec:mod}
In order to gain more physical insight from the measured pulse profiles, we applied the relativistic light bending code developed by \citet{phd_falkner}, which was previously used by \citet{iwakiri_spectral_2019} to successfully model pulse profiles of 4U\,1626$-$67. In this approach we, for now, do not consider the evolution with energy and therefore focus our efforts on a narrow energy band with high count rate. We selected the \nicer pulse profiles in the 3.5--4.5\,keV energy band. To constrain geometrical parameters and break some of the degeneracies of the model, albeit at the cost of increased computation time, we simultaneously fitted two pulse profiles at different luminosities with the same shared column position. We chose to fit the pulse profiles of the two \nicer observations that were concurrent with the two \nustar observations. Expanding the analysis to a wider range of energies and times will be part of future work.

The applied model is set up as follows: Sources of emission are two cylindrical accretion columns with height $h$ and radius $r$ placed at arbitrary positions on the neutron star surface. Each column consists of two types of emitting surface elements - those that make up the walls and those that make up the top of each column. For both the top and the wall surfaces, we define an emission pattern that is equal for all surface elements of one category and shared by both columns. 
While the column positions are constant between the two observations and assumed not to vary, the emission parameters are allowed to change between observations of varying luminosity.

This leads to four independent emission components, as we have two observations with top- and wall-emissions each. These four total emission components are each defined by a simple Gaussian emission profile. For the wall component, this Gaussian profile is perpendicular to the column axis and is therefore defined by its width $\sigma_\mathrm{wall}$. This is intended to model any kind of ``fan'' emission pattern. The emission from the top can be inclined from the column axis. Therefore, its Gaussian emission profile is centered at an angle $\mu_\mathrm{cap}$ and has a width of $\sigma_\mathrm{cap}$. This can mimic the butterfly pattern of a ``pencil'' beam predicted by \citet{basko_radiative_1975} among others.

\begin{figure}         
        \resizebox{\hsize}{!}{\includegraphics{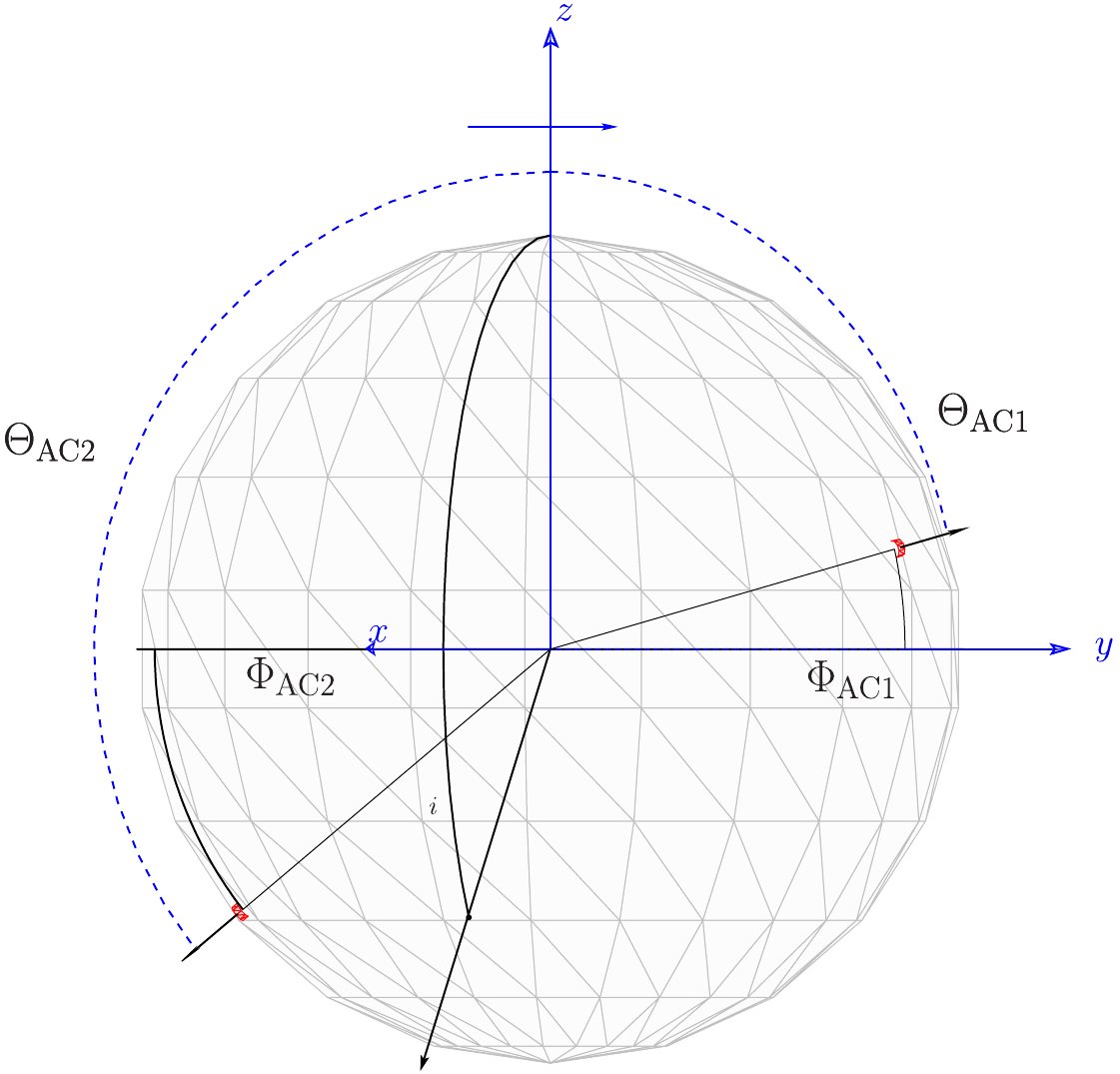}}
        \caption{Geometry of the two accretion columns (red), the rotation axis in the z-direction, and the coordinate system spanned by it (blue).}
        \label{fig:geom}
\end{figure}

The location of the two columns is determined by two angles: the polar angle $\Theta$, defined as the angle between the column and the rotation axis of the neutron star and the azimuthal angle $\phi$. Figure\,\ref{fig:geom} provides a depiction of the neutron star with two accretion columns and characteristic angles. We also have to set the observer's viewing angle $i$ with respect to the rotation axis.
In order to reduce the number of free parameters, we fixed the radius of each column at 250\,m and the height at 300\,m for both observations.
We keep these values fixed between the two observations as their influence on the pulse profiles is minor for most geometries. 

Still, we want to note at this point that the parameters of the accretion column and its exact geometry (a hollow or filled cylinder) are widely discussed in the literature, without a current consensus on the topic. We admit, however, that both the radius and the height of the column depend on the strength of the magnetic field, which in the first approximation is typically considered to be dipolar. The radius of an accretion channel near the surface of a highly magnetized accreting neutron star is expected to be small compared to the total surface area. Very approximately, it can be estimated from the magnetospheric radius, $R_\mathrm{m}$, and the neutron star radius, $R_\mathrm{NS}$, as $r_0\sim R_\mathrm{NS}\sqrt{R_\mathrm{NS}/R_\mathrm{m}}$ \citep{lamb_1973}. In the case of \exo, the magnetic field remains unknown, although a possibility of a shallow absorption line in the spectra was discussed at energies ${\sim}10$--$60$\,keV \citep[][and Paper~II]{klochkov_integral_2007}, this feature has not been confirmed as a cyclotron line so far. Alternative methods of field estimates provide a rather broad range of values, $B\sim0.5\times10^{11}$--$10^{13}\,\mathrm{G}$ \citep{tamang_spectral_2022}. It translates to the column radius of ${\sim}200$--$800\,\mathrm{m}$ for the accretion luminosity of ${\sim}10^{36}\,\mathrm{erg}\,\mathrm{s}^{-1}$ and canonical neutron star parameters (it also depends on the parameters of the accretion flow above the magnetospheric radius, which remain largely unknown as well; see, e.g., Eq.~18 of \citealt{lamb_1973} and Eq.~23 of \citealt{becker_2012}).

The height of the accretion column is even more uncertain. At luminosities above the critical one, it is expected that the radiation pressure is capable of initiating deceleration above the surface of the neutron star \citep{basko_limiting_1976}. The heights of the accretion columns typically discussed in one-dimensional analytical modeling are ${\sim}10\,\mathrm{km}$ \citep[see, e.g.,][]{basko_limiting_1976, becker_2012,West_2017}. At lower, intermediate luminosities, the estimate of the accretion column height varies, spanning a few orders of magnitude. Thus, it was discussed by different authors that the emission is either produced directly at the polar caps of the neutron star \citep{mushtukov_positive_2015}, slightly above the surface, by ${\sim}100$--$200\,\mathrm{m}$, where matter is being decelerated by Coulomb collisions \citep{staubert_2007} or a collisionless shock \citep{Vybornov_2017}, or kilometers above the surface in the frame of a different collisionless shock model \citep{becker_generalized_2022}. Some analytical estimates \citep{becker_2012, West_2017} heavily rely on estimates of the magnetic field and mass accretion rate, both of which are challenging for \exo due to an unconfirmed cyclotron line and uncertainty on the distance (see Sect. \ref{sec:dis}). In the choice of both the height and the radius of the accretion column, we relied on the most recent two-dimensional magnetohydrodynamic simulations by \citet{Sheng_2023}, which demonstrated a rather low height of the accretion column even in the case of sufficiently high radiation pressure. It is important to note, however, that within the frame of the model described above, the height of the column does play a crucial role in the variation of pulse profiles.

Recently, \textit{IXPE} observed \exo during one of its recent Type~I outbursts \citep{Malacaria_2023}. Importantly, the evolution of the polarization angle with pulse phase was modeled by the rotating vector model (RVM), which depends on the orientation of the magnetic dipole field. Thus, the pulsar inclination was constrained to $127.\!\!^\circ5^{+9}_{-7}$. Therefore, we adopt a value of 130\,$^\circ$ as the inclination in our model, as otherwise we could only rule out low inclination angles $\lesssim 40^\circ$, as seen in Fig.~\ref{fig:inlc}. The reliability of the RVM will be discussed toward the end of this paper.

\begin{figure}
   \resizebox{\hsize}{!}{\includegraphics{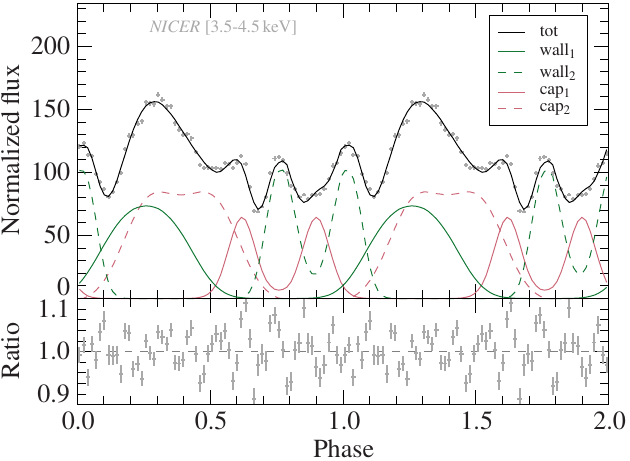}}
   \resizebox{\hsize}{!}{\includegraphics{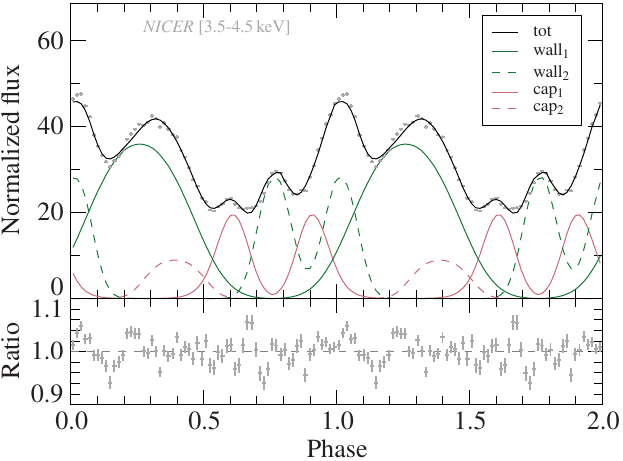}}
   \caption{Pulse profile of the high (top) and low (bottom) luminosity observation. Model components from the two columns with top (dashed) and wall (solid) emission are shown in red and green.}
   \label{fig:profsim}
\end{figure}

\begin{figure}
\resizebox{\hsize}{!}{\includegraphics{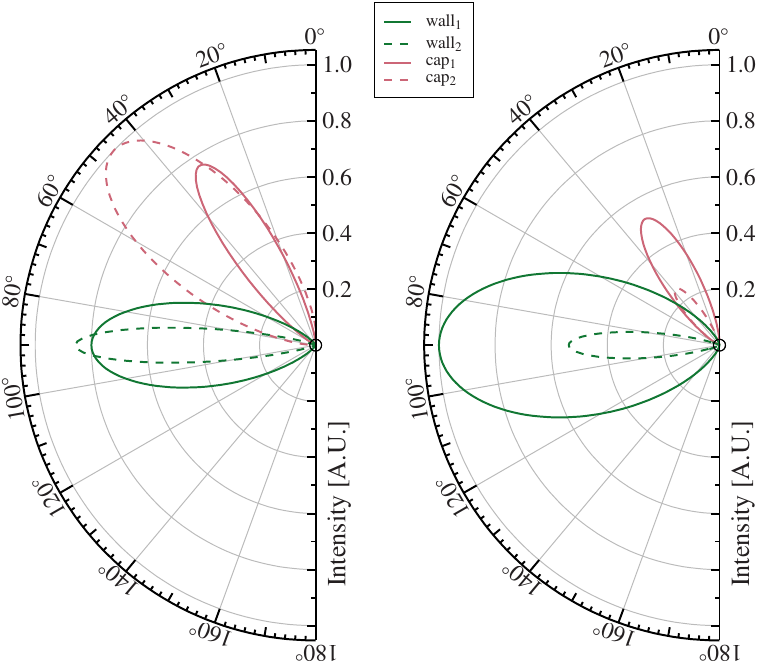}}
   \caption{Emission profiles of the different emission components giving rise to the profiles shown in Fig.~\ref{fig:profsim}. The left and right figure show the emission profiles for the high and low luminosity state respectively. The radial coordinate represents the integrated flux in the reference frame of the neutron star. }
   \label{fig:emprof}
\end{figure}

Then finally, to calculate the observed flux for each rotation phase from the setup just described, our model projects the surface elements onto the plane of the observer, while taking relativistic light bending into account, calculates their visibility and brightness according to their emission pattern, and integrates over them. This results in a model flux for each phase bin that can then be compared with the observed data. A more detailed description of the required calculations can be found in Appendix \ref{app:fancil}, in the application by \citet{iwakiri_spectral_2019}, and in 
 \citet{phd_falkner}.
We find the configuration described by Table\,\ref{tab:fancil}, with pulse and emission profiles of Fig.~\ref{fig:profsim} and Fig.~\ref{fig:emprof}, to give a good description of the observed profiles.

Trying to extend this model to profiles corresponding to even lower luminosities by simply adjusting the normalization of the four components only lead to unsatisfactory fits, confirming that not only the strength of cap or column emission changes, but the respective emission pattern of each component as well.

The geometry found is marked by an offset between the two columns of ${\sim}169^\circ$, deviating from a simple dipole geometry with two opposing columns. This is not unexpected as this is the only way to obtain the observed asymmetry of the profiles from cylindrically symmetric emission profiles in this model. An alternative would be contributions from more than two poles, which we would, however, not expect as the matter couples to the magnetic field at a radius where the dipole component dominates. %
While it would be possible to include more poles in our model, this would significantly increase its complexity and should only be done if required to describe the observed profiles, which is not the case for \exo.
Of note is the large uncertainty on the width of the second cap emission in the low-luminosity observation. This can indicate that the majority of this emission does not reach the observer and is therefore only weakly constrained by the data. The given statistical uncertainties depend on reliable fit convergence and are therefore likely underestimated.  
\begin{table}
  \caption{best fit model parameters for the high and low luminosity states.\label{tab:fancil}}
  \begin{tabular}{lrrr}
  \toprule
Parameter &  & $L_\mathrm{high}$ & $L_\mathrm{low}$ \\
\hline
$i_\mathrm{obs}$        & [deg] & \multicolumn{2}{c}{ 130  }$^\dagger$ \\    
col$_1$/col$_2$ ratio    &        &  $0.43^{+0.08}_{-0.07}$  &  $0.767^{+0.016}_{-0.031}$ \\  
\hline \\
cap$_1$/wall$_1$ ratio     &        &  $0.23^{+0.06}_{-0.12}$  &  $0.118^{+0.025}_{-0.013}$ \\    
$i_\mathrm{mag,1}$      & [deg] &  \multicolumn{2}{c}{ $116.^{+26.}_{-11.}$ }$^\dagger$ \\    
$\phi_\mathrm{1}$       & [deg] &  \multicolumn{2}{c}{ $324.1^{+1.1}_{-2.0}$ }$^\dagger$ \\    
$\sigma_\mathrm{col,1}$ & [deg] &  $18.9^{+2.2}_{-2.0}$  &  $27.1^{+3.6}_{-2.4}$ \\    
$\mu_\mathrm{cap,1}$    & [deg] &  $34^{+4}_{-8}$  &  $32.1^{+2.1}_{-1.7}$ \\    
$\sigma_\mathrm{cap,1}$ & [deg] &  $10.4^{+1.9}_{-1.5}$  &  $12.0^{+1.8}_{-1.4}$ \\    
\hline \\

cap$_2$/wall$_2$ ratio     &        &  $0.51\pm0.04$  &  $0.20^{+0.22}_{-0.09}$ \\    
$i_\mathrm{mag,2}$      & [deg] &  \multicolumn{2}{c}{ $75.6\pm0.6$ }$^\dagger$ \\    
$\phi_\mathrm{2}$       & [deg] &  \multicolumn{2}{c}{ $191.3^{+0.9}_{-1.4}$ }$^\dagger$ \\    
$\sigma_\mathrm{col,2}$ & [deg] &  $6.9^{+0.8}_{-0.7}$  &  $8.4^{+1.0}_{-0.7}$ \\    
$\mu_\mathrm{cap,2}$    & [deg] &  $53.9^{+1.8}_{-2.5}$  &  $39^{+5}_{-40}$ \\    
$\sigma_\mathrm{cap,2}$ & [deg] &  $21^{+7}_{-4}$  &  $10^{+20}_{-5}$ \\

    \bottomrule
  \end{tabular}
  
      \tiny Note: $^\dagger$ tied together for both observations 
\end{table}

\subsection{Empirical pulse profile fitting}  \label{sec:fit}
\begin{figure}
  \resizebox{\hsize}{!}{\includegraphics{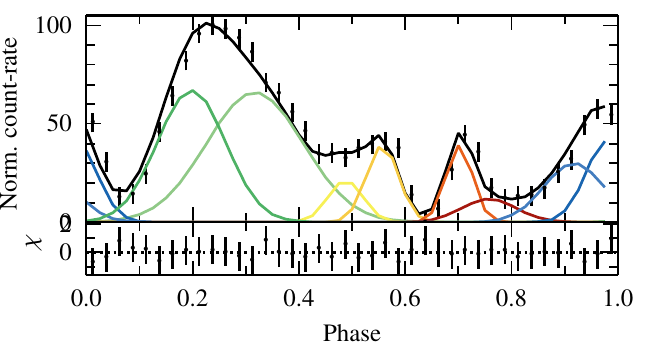}}
  \resizebox{\hsize}{!}{\includegraphics{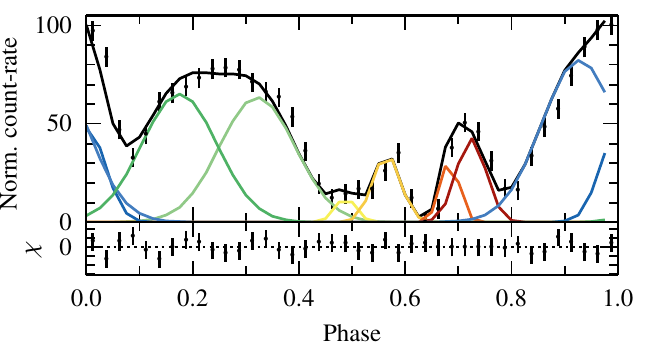}}
  \caption{Gaussian fits to observation 4201960113   (top) and 4201960150 (bottom) as an illustration of the phenomenological fitting approach. The black curve indicates the full model while the individual Gaussian peaks are shown in the same colors as in Fig.~\ref{fig:ppfitpar}}\label{fig:ppfitgauss}
\end{figure}

The evolution of the relative intensities of the individual peaks in the pulse profile, visible in Fig.~\ref{fig:map}, led us to investigate the evolution of the pulse profile over the full luminosity range. Due to the complexity of the physical model presented in the previous chapter and the large amount of pulse profile data available through \nicer, only a simple empirical model was applied to the profiles of all observations. We settled on the sum of several Gaussian emission features as a simple model to quantify the profile evolution and again apply this model to the 3.5--4.5\,keV energy band.
By visual inspection we find that six Gaussian components reproduce the basic pulse profile morphology; however, to reliably model the profiles for all luminosities, eight Gaussians were necessary (see Fig.~\ref{fig:ppfitgauss}). As each Gaussian is functionally identical, we restrict the position of each of them to a certain phase range to avoid components switching positions, thus hampering convergence of the applied fitting algorithm.  

\begin{figure}
  \resizebox{\hsize}{!}{\includegraphics{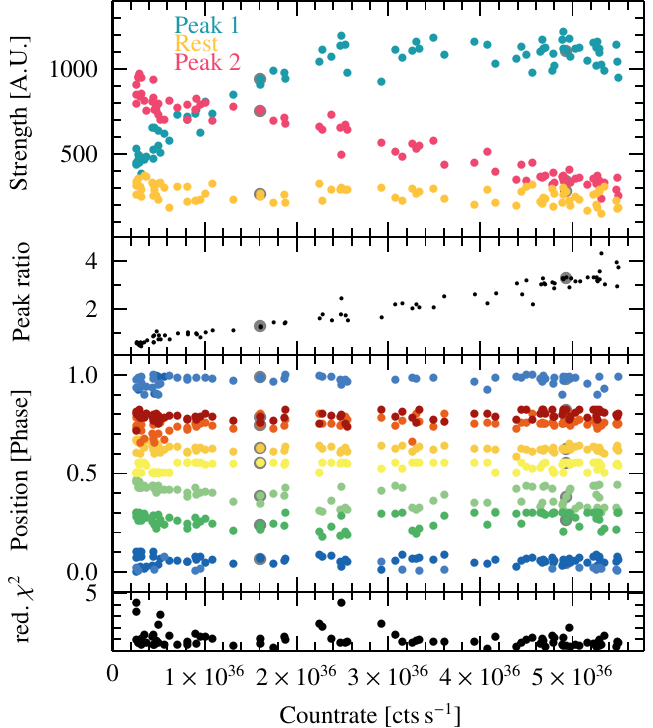}}
  \caption{Results of fitting Gaussians to the \nicer pulse profiles. Top: Strength of the two main peaks (red and blue) and secondary features (yellow). The first peak is defined as the sum of the first and last Gaussian, the second one as the second and third Gaussian. All other Gaussians are summed as secondary features. Middle: Ratio between the strength of the secondary and primary peak. Bottom: Position of all fitted Gaussians. The two observations with ObsID  4201960113 and 4201960150, selected also for Fig.~\ref{fig:ppfitgauss}, are marked by larger gray rings. 
 }\label{fig:ppfitpar}
\end{figure}

\begin{figure}
  \resizebox{\hsize}{!}{\includegraphics{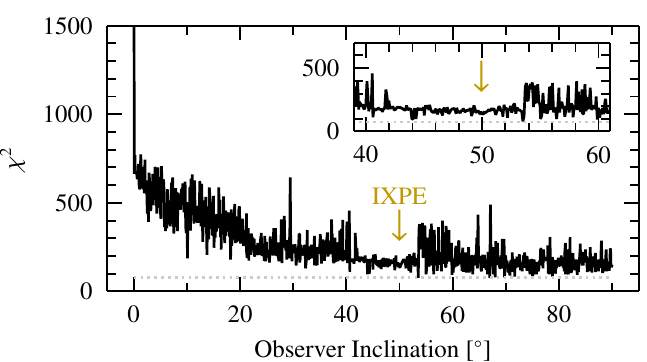}}
  \caption{Goodness of the best fit of our physical pulse profile model for a range of fixed observer inclinations. Marked is the inclination corresponding  derived from \ixpe data and used for our best fit \citep{Malacaria_2023}. Additionally the overall best $\chi^2$ value is indicated by the dotted gray line. The inlet focuses on the range around the \ixpe value.}\label{fig:inlc}
\end{figure}

The absolute flux evolution is already reflected by the light curve, so we fit only the relative strengths of the emission components and possible phase shifts. Using systematic uncertainties of five percent on the count rate leads to flat residuals around one for most observations. 
An overview of the fit results is shown in Fig.~\ref{fig:ppfitpar}. We see the second peak rapidly gaining strength toward higher luminosity, reaching a plateau around ${\sim} 3\times10^{36}\mathrm{erg\,s}^{-1}$ or a \nicer count rate of $300\,\mathrm{counts}\,\mathrm{s}^{-1}$. The first peak, on the other hand, more steadily loses its dominance toward higher luminosity, whereas the remaining features stay mostly constant. This behavior is consistent with that seen in Fig.~\ref{fig:map}.  

In addition to the description with Gaussian components, we followed the procedure of \citet{alonso-hernandez_common_2022} and separated the profiles into its sinusoidal Fourier components. \citet{alonso-hernandez_common_2022} identify pattern in the energy dependencies of the first three Fourier components to categorize X-ray pulsars into one of three types. For both \nustar observations \exo closely resembles sources of their Type~2, such as Cen~X--3. However, we found that the change in the profiles in \exo with luminosity results only in shifts in the phase offset of the Fourier components, whereas their respective amplitudes changed little. This is a consequence of the fact that individual Fourier components do not correspond to individual peaks in the profile, but rather model the profile as a whole. From this we conclude that the classification of \citet{alonso-hernandez_common_2022}, which only takes into account the amplitude of these features not their position, is insufficient to describe such variability as seen in \exo.  For this reason, we will not discuss such a decomposition into Fourier components in more detail here.

\section{Discussion and Conclusions}\label{sec:dis}

 In this paper, we have shown that the evolution of the spectrum, specifically the hardness ratio, over the course of this outburst follows the same luminosity dependence as seen in previous outbursts \citep{epili_decade_2017,reig_x-ray_1999}.
The observed softening with higher luminosity is seen in many BeXRBs \citep{reig_patterns_2013} and is interpreted as a sign that the source has exceeded its critical luminosity. A proposed explanation is the increased emission height, where photons can escape the accretion column with increasing luminosity, as this would reduce the efficiency of bulk Comptonization \citep{becker_2012}. Although this would contradict the assumption of a constant column height made while modeling the pulse profiles, we have seen that the emission height has only very little effect on the observed profiles and would primarily affect the spectral formation.  Alternatively a reduced fraction of photons interacting with the stellar surface as the emission regions increase with increases in luminosity \citep{postnov_dependence_2015}, can lead to the observed softening.

Similarly, the evolution of the pulse profile with luminosity fits remarkably well to what has been seen in previous Type~I and Type~II outbursts \citep{parmar_transient_1989-1,klochkov_giant_2008,naik_suzaku_2015,epili_decade_2017}.

Even the sharp dip in the pulse profile, with accompanying change in the hardness ratio, studied in detail by \citet{ferrigno_glancing_2016} reappears again at low luminosities. This feature was already seen with \rxte, which provided the necessary time resolution and effective area to resolve it \citep{reig_timing_1998}. In subsequent observations, this dip was attributed to obscuration of the emission region by a phase-locked accretion stream \citep{jaisawal_detection_2021,ferrigno_glancing_2016}. As the accretion geometry is determined largely by the geometry of the magnetic field, this indicates a very stable $B$-field configuration over almost four decades, unaffected by repeated periods of accretion and quiescence.

Several attempts have been made at modeling the pulse profiles and reconstruction of the underlying emission geometry of accreting magnetized neutron stars \citep{bulik_asymmetric_fits_1995,Kraus_1996,sasaki_analyzing_2010,Caballero_A0535+26_2011,sasaki_analyzing_2012,Cappallo_SXP348_2019,Cappallo_SXP1062_2020,Roy_SMCX2_2022}. Roughly two approaches can be distinguished: forward modeling of detected pulse profiles, and decomposition of individual emission components from the measured profiles, each requiring varying assumptions on geometry and emission profile. Applications to \exo are limited to early modeling work by \citet{parmar_transient_1989-1}, and pulse profile decomposition by \citet{sasaki_analyzing_2010}. 

The forward modeling approach of \citet{parmar_transient_1989-1} is similar to our work as it models the emission from two non-antipodally positioned accretion columns emitting both wall and cap components. Their emission functions are slightly different from ours, most important by using cap emission that is peaked in the direction of the magnetic field. 
Most importantly, however, light bending was only included in a very simplified manner and only a rough solution fitting some of the observations was found by eye. Even more severely than in our case, this limits the degree to which their solution can be relied on to be unique and physically accurate.

Applying our light-bending model to the pulse profiles showed that the profiles can be well described by a composition wall and cap emission originating from two accretion columns. As such, we can explain the observed complexity of the profiles at different luminosities solely through a simple emission profile and column geometry without the need of a dip induced by absorption. An exception is formed by the sharp absorption dip mentioned in Sect.~\ref{sec:dip}.
In our model, the asymmetry of the pulse profiles is explained by an offset of the two columns from a simple dipole geometry, similar to previous work \citep{parmar_transient_1989-1,sasaki_analyzing_2010}. We found that the angle subtended by the two column positions is $119^\circ$ in our model, compared with the $110^\circ$  estimate by \citet{parmar_transient_1989-1}, $140^\circ$ estimated by \citet{sasaki_analyzing_2010}. \footnote{For directly opposed columns in a simple dipole field this angle would be $180^\circ$.}
These variations should not be surprising, considering the simplified relativistic treatment of \citet{parmar_transient_1989-1} and the entirely different approach of \citet{sasaki_analyzing_2010}. Their decomposition of the measured profiles and reconstruction of the local emission pattern relied on the assumption of cylindrical symmetric and identical emission from two columns. This technique was first introduced by \citep{Kraus_1996} and was since then applied to multiple sources. Its assumption of equal emission from the two columns fundamentally contradicts the emission profiles found in our work, and we were unable to reproduce the data when fixing the emission profiles of the two columns together.

Contrary to theoretical expectations \citep{davidson_1973,basko_limiting_1976}, the emission pattern that we found to reproduce the observed pulse profiles displays a stronger top or ``pencil'' emission for the high-luminosity observation. This could be explained by reflection of downward-boosted wall emission from the neutron star surface. This component could then appear as cap emission in a model such as ours \citep{poutanen2013,postnov_dependence_2015}. 
 We rely on the observer inclination determined through the application of the RVM to recent \textit{IXPE} observations of  \exo  to fix this parameter in our pulse profile model \citep{Malacaria_2023}.
 \textit{IXPE} observations of the source LS~V\,+44\,17 have, however, shown that a simple application of the RVM, can lead to misleading geometries when emission components with constant polarization signal with pulse phase are ignored \citep{Doroshenko_2023}. The origin of such a constant component is suspected to be reflection off of material in the disk or an outflow close to the neutron star. However, since there have been cases in which the inclination found with the RVM seems to be reliable, as it aligns well with the orbital inclination \citep{ Suleimanov_2023,Mushtukov2023} and observations that do not require a phase-independent reflection component despite multiple \ixpe observations \citep{Tsygankov_2023}, we still consider the inclination found by \citet{Malacaria_2023}, to be reliable enough to use it as a basis for our pulse profile modeling.  
 Multiple \textit{IXPE} observations would still help to resolve such ambiguities, but might only be possible during the next Type~II outburst.
Interestingly, the assumed inclination of $50^\circ$ in the decomposition of the emission pattern by \citet{sasaki_analyzing_2010} fits well with the RVM results, taking into account the $180^\circ$ symmetry of the problem. 

Pulse phase-resolved spectroscopy has shown a significant phase dependence of the column density \citet{reig_timing_1998,naik_suzaku_2015,naik_timing_2013,Malacaria_2023}, which is currently not taken into account in our model, since only Gaussian emission profiles are assumed. Still, it should be noted that multiple ``dips'' in the profiles can be reproduced just from the combination of simple emission profiles and light bending, without relying on absorption.
The important next step in our investigation of the pulse profiles of \exo is a phase-resolved spectral analysis of the combined \nicer and \nustar data, which is presented in Paper~II.

Further interpretation of the emission profiles and column geometry found should be done with care, as we cannot guarantee that we have found the unique physical solution capable of describing the measured pulse profiles.  One way to verify the uniqueness of our solution would be to run algorithms that are more suited for sampling a wider range of the parameter space, since the complex parameter landscape of the problem leads to a prohibitively long burn-in and autocorrelation time when attempting to a Markov chain Monte Carlo (MCMC) analysis.

An important next step is to constrain the column and emission parameters used in our model through the application of  self-consistent hydro dynamical models such as those of \citet{West_2017} to the phase-averaged and phase-resolved spectra of \exo. However, this is outside the scope of this work and will be part of forthcoming publications.

We can, therefore, say that it is possible to produce the complex pulse profiles of \exo and their evolution through simple Gaussian emission profiles that are emitted by two columns, which is far from obvious given the complex emission profiles shown in Fig.~6 in \citet{sasaki_analyzing_2012}. While, given the previous paragraph, we cannot yet be confident to have found the only solution that can describe the date, the fact that a satisfactory solution can be found with our model is already a major step toward understanding the geometry of and emissions from accreting neutron star pulsars.

Over the course of the outburst, we observed two types of transitions in the emission of \exo: a transition in the pulse profiles and one in the hardness-luminosity relation. The spectral transition occurs, as is visible in Fig.~\ref{fig:hardness}, around $5.0\times10^{35}\, \mathrm{erg}\,\mathrm{s}^{-1}$.
The transition in the profiles appears to occur around a significantly higher luminosity of $1.7\times10^{36}\, \mathrm{erg}\,\mathrm{s}^{-1}$. Assuming these to be signatures of one single state transition, it appears to manifest itself first in the spectrum before becoming apparent in the pulse profiles.

From these estimates of a critical luminosity, $L_\mathrm{crit}$, we can derive a corresponding $B$-field estimate and compare this with existing claims. There are two theoretical derivations of $L_\mathrm{crit}$, one analytical by \citet{becker_2012} and one numerical by \citet{Mushtukov_2015}, both of which we want to apply for this purpose. 
Their assumptions differ in that \citet{becker_2012} equates the critical luminosity with the Eddington luminosity, while \citet{Mushtukov_2015} requires the matter to come to rest at the surface. Further, \citet{becker_2012} approximates the cross section by assuming photon energies below the cyclotron energies, while \citet{Mushtukov_2015} considers the effects of cyclotron resonance.  The assumptions on the dominance of wind- or disk-accretion and the assumed spectral shape differ between the two works. For further details, consult the respective publications.

\citet{becker_2012} derives the critical luminosity as 
\begin{equation}
    1.5 \times 10^{37} B_{12}^{16/15},\mathrm{erg}\,\mathrm{s}^{-1},
\end{equation}
while \citet{Mushtukov_2015} present a numerical solution in their Fig.~7. They derived $L_\mathrm{crit}$  for the two cases of purely extraordinary, polarization and an equal fraction of ordinary and extraordinary polarization, with the reality mostly likely lying in-between.

Theoretical models usually calculate the full bolometric X-ray luminosity in the rest frame of the neutron star surface. In order to compare our measured luminosities with theory, we therefore fit an extended model, including a cut-off power law to the two \nustar observations and their corresponding \nicer data. For these, the 0.1--100\,keV flux is 2.74 and 2.79 times higher than the flux in the 2--10\,keV range. Therefore, we can multiply the previously stated observed transition luminosity by the average value of $2.75\times(1+z)^2$, also accounting for the gravitational redshift of $z=0.3$. This leads to a critical bolometric luminosity in the neutron star rest frame of $2.3\times10^{36}\,\mathrm{erg}\,\mathrm{s}^{-1}$ for the spectral transition and $7.9\times10^{36}\, \mathrm{erg}\,\mathrm{s}^{-1}$ for the transition seen in the pulse profiles.

Using these as a range for $L_\mathrm{crit}$ we can derive a surface magnetic field of 0.17--0.55\,B$_\mathrm{12}$ for the model by \citet{becker_2012}. According to  \citet{Mushtukov_2015}, a low $L_\mathrm{crit}$ can only be reached with mixed polarization and a surface $B$-field around 0.9--$1.7\times B_\mathrm{12}$, while an $L_\mathrm{crit}$ close to $7.9\times10^{36} \mathrm{erg}\,\mathrm{s}^{-1}$ would exclude this $B$-field range. Therefore, without further assumptions of the locally dominant polarization mode, no further restriction on the $B$-field strength can be made \footnote{\ixpe observations presented by \citet{Malacaria_2023} do not provide a reliable estimate of the dominant polarization mode due to uncertainties in the $B$-field geometry and poorly understood effects of the ``vacuum resonance'', by which photons can change polarization mode during propagation through the magnetosphere.}.

As this work focuses on the monitoring data obtained by \nicer, a very similar analysis of the Type~II outburst of 2021 was performed by \citet{Fu_2023} based on \hxmt , instead of \nicer data. They also provide several estimates of the magnetic field strength of \exo. First, by determining the critical luminosity from the pulse profile transition and, secondly, by applying the torque model by \citep{Ghosh_1979}. \citet{Fu_2023}, however, use distances of 3.6\,kpc and 7.1\,kpc in their analysis. As argued in Appendix~\ref{sec:distance}, we judge the updated distance of 2.4\,kpc as more reliable and therefore have to convert their reported values. 
The first method leads to a critical flux of $F^{\mathrm{B12}}_\mathrm{crit}{\sim}1.1\times 10^{-8}\,\mathrm{erg}\,\mathrm{cm}^{-2}\,\mathrm{s}^{-1}$, or $L^{\mathrm{B12}}_\mathrm{crit}{\sim}1.7\times 10^{37}\,\mathrm{erg}\,\mathrm{s}^{-1}$ for a distance of 2.4\,kpc and after correcting for gravitational redshift. This value is slightly higher than the critical luminosity determined similarly from the \nicer data, probably due to the different energy ra-nge used. 
From this a $B$-field of ${\sim}0.5\times10^{12}\,\mathrm{G}$ can be derived following \citet{becker_2012}. 

Alternatively, \citet{Fu_2023} derived a $B$-field strength through torque modeling. To that end, they fitted the proportionality constant between the spin-frequency derivative $\dot{f}$ and $F^{6/7}$, where $F$ is the source flux, measured by \hxmt. This constant can be compared with an expression depending on  the $B$-field strength and the distance $D$, derived from the theory of \citet{Ghosh_1979}. In this way \citet{Fu_2023} arrived at their Eq.~7, deriving the $B$-field strength as a linear function of  $D^{-6}$. 
They used this relation to estimate a $B$-field strength of ${\sim} 0.41\,\times\,10^{12}\,\mathrm{G}$ and ${\sim} 24\,\times\,10^{12}\,\mathrm{G}$ for a distance of 7.1\,kpc and 3.6\,kpc, respectively. We can apply the same equation to the distance of 2.4\,kpc, as generally assumed in our work, leading to a  $B$-field strength of ${\sim}270\times\,10^{12}\,\mathrm{G}$. This exceedingly high value is due to the strong $D^{-6}$ dependence of the magnetic field strength.
However, it is unclear whether, in the fit of \citet{Fu_2023}, relativistic corrections were applied to calculate the intrinsic luminosity in the neutron star rest frame. From Eq.~5 in \citet{Fu_2023} one can see that $B\,\propto\, L^{-3}$. As the intrinsic luminosity is higher by a factor of $(1+z)^2$, this correction alone leads to a magnetic field strength that is ${\sim}5$ times lower. Similarity, any over- or underestimation of the intrinsic luminosity due to beaming effects, which can be on the order of a factor 4--10 \citep{phd_falkner}, will have strong effects on the derived magnetic field strength. 

\citet{Fu_2023} further point out that torque modeling and the critical luminosity sample the magnetic field at different radii from the neutron star and a deviation can be seen as a sign of a multipolar magnetic field. However, it should be noted that since \citet{Ghosh_1979} there have been more approaches toward torque modeling with varying assumptions on the disk-magnetosphere interaction, which does have an effect on the estimated magnetic field \citep[see, e.g., ][]{wang_1995,dai2006,rappaport2004a}. Together, these effects lead to very large systematic uncertainties on the $B$-field estimates by \citet{Fu_2023}, which prevents us from claiming that these values would contradict each other.

Finally, we note the significantly lower peak luminosity reached during this latest outburst compared to the two previously observed Type~II outbursts \citep{epili_decade_2017,klochkov_giant_2008}. This might correlate with the earlier onset of the outburst than predicted by \citet{laplace_possible_2017}, leading to a potentially less saturated Be-decretion disk.

In summary, using both \nicer and \nustar data covering the third detected giant outburst of \exo, we investigated its changing spectral and timing behavior over the course of the outburst. 
We detected the familiar softening of the spectrum with increasing luminosity that fits well with previous outbursts \citep{parmar_transient_1989,reig_patterns_2013}. We describe the pulse profiles through phenomenological fits to quantify the observed changes in the profile and successfully applied a physical light bending model to the profiles at high and low accretion rates. The multiple peaks and dips of the profile can be described simply as a result of a two-component (wall and top) emission pattern originating from two accretion columns. No further components from a more complex accretion geometry or absorption features are required. 
To further constrain the geometry with this approach, we would need to sample the entire parameter space and determine the uniqueness of the solution found. Methods such as nested sampling would provide a possible way to achieve this, but are beyond the scope of this paper. 

\begin{acknowledgements}
The authors thank the \nustar and \nicer teams
for approving our DDT requests and scheduling this dense monitoring campaign, and the XMAG Collaboration for their helpful comments and fruitful discussions. RB acknowledges support by NASA under award number 80NSSC22K0122. The material is based upon work supported by NASA under award number 80GSFC21M0002. ESL\ and JW\ acknowledge partial funding under Deutsche Forschungsgemeinschaft grant WI 1860/11-2 and Deutsches Zentrum f\"ur Luft- und Raumfahrt grant 50 QR 2202. We acknowledge funding from the ESA faculty on work on neutron star pulse profiles. This work has made use of data from the European Space Agency (ESA) mission
\textit{Gaia} (\url{https://www.cosmos.esa.int/gaia}), processed by the \textit{Gaia}
Data Processing and Analysis Consortium (DPAC,
\url{https://www.cosmos.esa.int/web/gaia/dpac/consortium}). Funding for the DPAC
has been provided by national institutions, in particular the institutions
participating in the \textit{Gaia} Multilateral Agreement.
\end{acknowledgements}

\bibpunct{(}{)}{;}{a}{}{,} 
\bibliographystyle{bibtex/aa} 
\bibliography{bibtex/literature} 

\appendix

\graphicspath{{fig/}}
\def\arraystretch{1.5}

\section{Column emission model}\label{app:fancil}

One of the proposed mechanisms that form characteristic pulse profiles
(PP) of accreting neutron stars is the formation of accretion columns
at the magnetic poles. These columns are believed to form due to an
accretion shock \citep{becker_2012} and result in a cone-like structure
that potentially emits photons from its walls and cap. To model phase-dependent emission we propose a simplified model for this
emitting surfaces based on the work of \citet{phd_falkner}.

\subsection{Solution to emitting surfaces problem in Schwarzschild metric}

To calculate the total flux as seen by an observer, it is necessary to
determine the visible surface of the emitting surface elements taking
any blocking surface into account. \citet{phd_falkner} proposed an
approximation by describing the NS surface and column surface by a
mesh grid and project any of the mesh vertices into the observer sky
taking into account photon trajectory solutions of the Schwarzschild
metric \citep{misner_1973}. The visible surface of the columns is then
given by all surface elements that are not behind any other surface
element. This is a computational effort and is simplified by
considering the centroid of any given surface element only. If this
centroid is within the (projected) boundaries of another surface
element in front of it, it is deemed not visible.

Additional edge cases are taken into account as described in
\citet{phd_falkner}.

For simplicity, we approximate the column structure as a cylinder that
emits from the wall with surface $A_\textnormal{wall} = \pi \rho h$,
with the column radius $\rho$ and column height $h$, and from the cap
with emitting surface $A_\textnormal{cap} = 2\pi \rho^2$. Each surface
is divided into a fine grid of surface elements where each element is
assumed to emit with an intensity in a given direction according to a
Gaussian distribution centered around the surface normal. For the
emission from the cap, we additionally allow for an offset from the
surface normal such that we can account for potential suppression of
the cap emission due to infalling matter.

\subsection{Table model interpolation}

The pulse profile is modeled assuming a NS with canonical mass $M =
1.4\,\textnormal{M}_\odot$ and radius $R = 10^6$\,cm. Considering a
(distorted) dipolar field, we place two emission columns on the surface
with the location given by the angular displacement $\Theta$ from the
rotation axis and a phase shift $\textnormal{d}\phi$ relative to the
line of sight direction (such that $\textnormal{d}\phi=0$\,deg corresponds
to the column being in the direction of the observer). The NS
orientation is given by the inclination of the rotation axis $i$. For
$i=0$\,deg the rotation axis points toward the observer. All parameters
for one column are summarized in Table~\ref{tab:parameters}.

\begin{table}
\caption{Summary of model parameter}
\begin{tabular*}{\hsize}{@{\extracolsep{\fill}}rlp{126pt}}
Parameter & Range & Description \\
\hline
\multicolumn{3}{l}{\texttt{column}} \\
\hline
$\theta$ & 0--$180^\circ$ & Angular distance of magnetic pole to rotation
axis \\
$i$ & 0--$180^\circ$ & Inclination of rotation axis to observer
direction \\
$h$ & 1--5\,000\,m & Height of accretion column \\
$\rho$ & 1--1\,000\,m & Radius of accretion column \\
d$\phi$ & 0--$360^\circ$ & Phase shift (relocates points where pole is
in plane of rotation axis and observer direction) \\
\hline
\multicolumn{3}{l}{\texttt{wall}/\texttt{cap}} \\
\hline
$N$ & 0--$\infty$ & Total flux \\
$\mu$ & 0--$90^\circ$ & Direction of principle emission (only for
\texttt{cap}) \\
$\sigma$ & $10^{-6}$--$\infty$ & Width of emission beam \\
\hline
\end{tabular*}
\label{tab:parameters}
\end{table}

Placing two columns on the NS surface, we want to find the total
emission reaching the observer for any given rotational phase $\phi$.
Due to the computational effort, using the described method directly as
a model is tedious. Instead, one can calculate an interpolation table
for all parameters. While this is certainly possible, it has the
disadvantage that at least 19 parameters are necessary, resulting in
large tables and also slow interpolation. Additionally, if one wants to
change the emission profiles, recalculation of the tables is required.

Instead, we can improve the interpolation model drastically under the
following assumptions:
\begin{itemize}
\item Emission of one column is not blocked by the other.
\item Emission profile of all surface elements of cap and wall, respectively of one column are the same.
\item Emission profiles of the surface elements are radially symmetric
(with respect to the surface normal).
\item Columns are radially symmetric around the
$\Theta,\text{d}\phi$ direction.
\item Time delay effects are negligible.
\end{itemize}
With this, to calculate the total flux for a given rotational phase it
suffices to know the total visible surface $A(\gamma)$ observed under
a certain angle $\gamma$ for the cap and wall of a column separately.
Given this information one can integrate over the product of emission
function $f$ and surface $A$ for all angles to get the total emission
\begin{equation}
\begin{split}
F = \int_0^\frac{\pi}{2} \textnormal{d}\!\gamma & f_\textnormal{wall}(\gamma; \bar{\lambda})
A_\textnormal{wall}(\gamma; h, \rho, \Theta) \\
& + f_\textnormal{cap}(\gamma; \bar{\lambda})
A_\textnormal{cap}(\gamma; h, \rho, \Theta).
\end{split}
\end{equation}
Here, $\bar{\lambda}$ are additional parameters for the emission
profile $f$. As before we calculate a table model, now for
$A_\textnormal{wall}$ and $A_\textnormal{cap}$ where only three parameters
are required. Due to symmetry it suffices to consider only one angular
displacement parameter (here $\Theta$), arbitrary orientations are
obtained by rotation. With this, not only the resulting table is much
smaller, but also the total number of calculations necessary to
populate the table are reduced such that changing the model is less
time consuming.

\subsection{Fancil model}

With the previously described strategy we rebuild the fancil model
\citep{phd_falkner,iwakiri_spectral_2019} where $f$ is a Gaussian distribution,
centered around $\gamma = 0$\,rad for the wall emission, and centered
around $\gamma = \mu$ for the cap emission. A large width of the
Gaussian distribution corresponds to isotropic emission from each
element.

To implement the fancil model we make use of the programmatic access
of the internal data and fitting process in \emph{ISIS}
\citep{ISIS}. Here we define two functions \texttt{wall} and
\texttt{cap} which implement the emission profile. Due to the made
assumptions these profiles are functions of one variable only and can
in principle be replaced with other potential patterns. A third
function \texttt{column} takes the output of the previous two
functions, interpolates the calculated table and performs the
integration for all observed rotational phases.
The result is the pulse profile for the given set of parameters.

\subsection{Limitations}

It is clear that due to the made assumptions the model is severely
limited in some ways. The potential largest impact comes from the
assumed radial symmetry of emission. It is known \citep{Kraus_1996}
that emission from the wall is boosted toward the surface, hence, not
symmetric. Further, the assumption of symmetry around the magnetic axis ($\Theta$ direction) is likely not reflecting the physical reality. Due to the fast rotation of the magnetic axis it is reasonable to believe that the accretion column is not symmetric around this axis. With this
approach it is in principle possible to model more complicated
multipole configurations by adding more columns. However, this results
in columns being close to each other such that the shadowing between
the columns can no longer be neglected. For this case it is
necessary to solve the full problem which requires better algorithms
to be applied as a fitting model. Besides the geometric effects we
ignore any time dilation effects resulting in the reddening of emitting
spectra. It is in principle possible to use the \texttt{column}
function to predict phase-dependent spectra, but energy shifting
effects need to be considered separately.

\section{Distance to \exo}\label{sec:distance}
\begin{figure}
  \resizebox{\hsize}{!}{\includegraphics{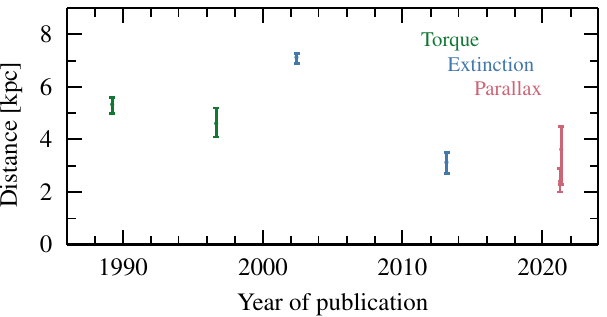}}
  \caption{Literature distances toward \exo. Included are \citet{parmar_transient_1989,Reynolds_1996,wilson_decade_2002,Coleiro_2013,Bailer-Jones_2021,Arnason2021}, each colored by the technique used to determine the distance.    }\label{fig:dis}
\end{figure}

Many of the astrophysical implications of observed state transitions depend on the luminosity at which the transition occurs, and therefore we have to establish the distance toward \exo. Previous distance measurements were derived from modeling the spin-up during Type~II outbursts \citep{parmar_transient_1989,Reynolds_1996}, optical extinction \citep{wilson_decade_2002,Coleiro_2013}, and from direct Gaia parallax measurements \citep{Gaia_2016,Gaia_2018} as derived by \citet{Bailer-Jones_2021}. See Fig.~\ref{fig:dis} for an overview of how the different methods yielded different distance estimates over time. Historically, a distance of 7.1\,kpc has been used by most publications since \citet{wilson_decade_2002}. As also discussed by \citet{Coleiro_2013}, \citeauthor{wilson_decade_2002} assumed that the extinction in the Galactic plane is representative for the optical extinction toward \exo. Using a more careful approach that takes into account multiple magnitude measurements in different optical bands to fit both extinction and distance, \citeauthor{Coleiro_2013} therefore revised the extinction-based distance down to $3.1\pm0.4$\,kpc.

This distance is consistent with direct parallax measurements from Gaia DR2 data, $3.6^{+0.9}_{-1.3}$\,kpc \citep{Arnason2021}, which in the more recent early EDR3 data release has been revised to $2.4^{+0.5}_{-0.4}$\,kpc \citep{Bailer-Jones_2021}. As the latter includes the most recent Gaia data, equivalent to DR3 \citep{Gaia_2023}, we judge this distance as most reliable and will use it in this paper to calculate luminosities, if not stated otherwise. As a consistency check,
for each distance in the literature, we determine the absolute magnitude of the optical companion calculated from de-reddened Gaia G-band magnitudes. We find that the distance of 2.4\,kpc results in a more realistic $M_\mathrm{V}=-0.98\,\mathrm{mag}$ for a B star, compared to $M_\mathrm{V}=1.52\,\mathrm{mag}$ for 7.1\,kpc.  
It should be noted that outside of the results presented here, the newer, lower Gaia distance also has significant implications for the interpretation of low luminosity observations, such as the detection of pulsations described by \citet{furst_studying_2017}. Recently \citet{Fu_2023}, very similar to our work, used \textit{Insight}-HXMT monitoring of the 2021 outburst to constrain the critical luminosity from a transition point in the pulse profiles and from that estimate the $B$-field. However, \citeauthor{Fu_2023} only consider distances of 7.1\,kpc \citep{wilson_decade_2002} and 3.6\,kpc \citep{Arnason2021}, leading them to estimate higher luminosities compared to this work.

\end{document}